\documentstyle[12pt]{article}

\newcommand{\bm}[1]{\mbox{\boldmath $#1$}}
\newcommand{\be}{\begin{equation}}
\newcommand{\ee}{\end{equation}}
\newcommand{\ba}{\begin{eqnarray}}
\newcommand{\ea}{\end{eqnarray}}
\newcommand{\lb}{\label}
\newcommand{\ds}{\displaystyle}
\newcommand{\ra}{\rightarrow}
\newcommand{\ol}{\overline}
\newcommand{\bb}[1]{\bibitem{#1}}
\begin{document}
\begin{titlepage}
\setcounter{page}{1}
\title{Topologically massive gravito-electrodynamics: exact solutions}
\author{Karim Ait Moussa\thanks{on leave of absence from D\'epartement de 
Physique Th\'eorique, Universit\'e de Constantine, Constantine, Alg\'erie} \\
\small and \\
G\'erard Cl\'ement\thanks{E-mail: gecl@ccr.jussieu.fr} \\
\small Laboratoire de Gravitation et Cosmologie Relativistes \\
\small Universit\'e Pierre et Marie Curie, CNRS/URA769 \\
\small Tour 22-12, Bo\^{\i}te 142, 4 place Jussieu, 75252 Paris cedex 05, 
France}
\bigskip
\date{February 16, 1996}
\maketitle
\begin{abstract}
We construct two classes of exact solutions to the field equations of
topologically massive electrodynamics coupled to topologically massive
gravity in 2 + 1 dimensions. The self-dual stationary solutions of the
first class are horizonless, asymptotic to the extreme BTZ black-hole
metric, and regular for a suitable parameter domain. The diagonal solutions
of the second class, which exist if the two Chern-Simons coupling
constants exactly balance, include anisotropic cosmologies and static
solutions with a pointlike horizon. 
\end{abstract}
\end{titlepage}

\section{Introduction}
Topologically massive gravity (TMG) \cite{DJT} is a theory of gravity in
three-dimensional spacetime which includes, in addition to the otherwise
dynamically trivial Einstein-Hilbert action, a gravitational Chern-Simons
term. The resulting theory has non trivial dynamics with massive
excitations. A number of papers have been devoted to the construction of
exact solutions to sourceless TMG \cite{Vuo}\cite{sols}, as well as to TMG
with a cosmological constant \cite{Nut}\cite{part}. On the other hand, not
much is known about exact solutions to the theory with matter sources,
except for the special cases of point particles with a particular value
for the mass-to-spin ratio \cite{point}, of lightlike particle sources
\cite{DeserSteif}, and of a two-fluid source \cite{Williams}. Of special
interest is the case of TMG coupled to a spin 1 Abelian gauge
(``electromagnetic'') field. In this case, it seems natural to also add to
the minimal Maxwell action a Chern-Simons term for the electromagnetic
field, leading to the theory of topologically massive electrodynamics
(TME) \cite{Sch}\cite{DJT}. Exact solutions of TME coupled to Einstein
gravity, generalizing solutions of Maxwell electrodynamics coupled to
Einstein gravity [10--12] have been discussed in \cite{Cam}. In this paper,
we shall discuss classical solutions of TME coupled to TMG ---
topologically massive gravito-electrodynamics (TMGE), and construct two
classes of exact solutions.

In the second section, we generalize to the case of the fully coupled TMGE 
theory with a cosmological constant the methods previously used to reduce, 
under the assumption of two commuting Killing vectors, the field equations 
of TMG \cite{part} or of TME coupled to Einstein gravity \cite{Cam} . A
first class of exact, self-dual stationary solutions are constructed in
the third section. These horizonless solutions, which exist only for a
negative or zero cosmological constant $\Lambda$, are for $\Lambda < 0$ 
asymptotic to the extreme BTZ black hole metric \cite{BTZ} . In the fourth
section we investigate the existence of diagonal solutions to TMGE. It is
known that there are no non-trivial static solutions (diagonal stationary
solutions) either to TMG \cite{Vuo} or to TME coupled to Einstein gravity
\cite{Cam}. However one may speculate that the dynamical spins generated
by the two Chern-Simons couplings could exactly balance so that spinless,
static solutions would be possible. Indeed we shall find exact diagonal
solutions for a particular relation between the two Chern-Simons coupling
constants. These solutions are cosmologies for $\Lambda > 0$, and static
solutions for $\Lambda < 0$. We summarize our results in the last section.

\setcounter{equation}{0}
\section{Reduction of the field equations}

The action for TMGE may be written
\be \lb{1}
I = I_E + I_M + I_{CSG} + I_{CSE}\,,
\ee
where 
\ba \lb{2}
I_E & = & -m \int d^3x \sqrt{|g|}\, (R+2\Lambda)\,,  \nonumber \\
I_M & = & -\frac{1}{4} \int d^3x \sqrt{|g|}\, g^{\mu\nu}g^{\rho\sigma}F_{\mu
\rho}F_{\nu\sigma} \,. 
\ea
are the Einstein action (with cosmological constant $\Lambda$ and
gravitational constant $G \equiv 1/16\pi m$) and the Maxwell action, and
\ba \lb{3}
I_{CSG} & = & -\frac{m}{2\mu_G} \int d^3x \,\epsilon^{\lambda\mu\nu}\,
\Gamma^\rho_{\lambda\sigma} \left[\partial_\mu
\Gamma_{\rho\nu}^\sigma+\frac{2}{3}\,\Gamma^\sigma_{\mu\tau}
\Gamma_{\nu\rho}^\tau \right]\,,  \nonumber \\ 
I_{CSE} & = &
\frac{\mu_E}{2} \int d^3x \,\epsilon^{\mu\nu\rho} A_\mu
\partial_\nu A_\rho\,,
\ea
are the gravitational and electromagnetic Chern-Simons terms (with
$\epsilon^{\mu\nu\rho}$ the antisymmetric symbol). The Chern-Simons terms
being only pseudo-invariant, the absolute signs of the Chern-Simons
coupling constants $\mu_G$ and $\mu_E$ are unimportant (however we expect
their relative sign to be of importance). In the
case of sourceless TMG ($\mu_E = 0$), the gravitational coupling constant
$m$ should be negative (instead of positive as in four dimensions) to
avoid the appearance of ghosts \cite{DJT}. While this argument carries over
to the case of the coupled theory ($\mu_E \neq 0$) treated perturbatively,
we cannot rule out the possibility that the choice $m > 0$ might lead to a
consistent non-perturbative quantum theory, so we will leave the choice of
the sign of $m$ open.

We assume that the spacetime has two commuting Killing vectors, and choose
the parametrisation \cite{cosm}\cite{EML} 
\be \lb{4}
ds^2=\lambda_{ab}(\rho)\,dx^a dx^b + \zeta^{-2}(\rho)R^{-2}(\rho) \,d\rho^2\,,
\qquad A_\mu \,dx^\mu = \psi_a(\rho) \,dx^a
\ee
($a,b=0,1$), where $\lambda$ is the $2 \times 2$ matrix
\be \lb{5}
\lambda = \left( 
\begin{array}{cc}
T+X & Y \\ 
Y & T-X
\end{array}
\right), 
\ee
with ${\rm det}\,\lambda = R^2 \equiv \bm{X}^2$, the Minkowski pseudo-norm 
of the ``vector'' $\bm{X}$ of components $X^0 \equiv T$, $X^1 \equiv X$,
$X^2 \equiv Y$:
\be
\bm{X}^2 = X^iX_i = T^2-X^2-Y^2 \,,
\ee
and the scale factor $\zeta$ allows for arbitrary reparametrizations
of the variable $\rho$. We shall discuss in this paper both the cases of
solutions with $\bm{X}$ ``spacelike'' ($R^2 < 0$), for which the metric
(\ref{4}) is Lorentzian, $\rho$ being the radial coordinate, or
``timelike'' ($R^2 > 0$), corresponding to a Riemannian metric for $T >
0$, and to a cosmology for $ T < 0$ ($\rho$ is then the time coordinate). 

The parametrization (\ref{4}) reduces the action (\ref{1}) to the form
\be \lb{7}
I = \int d^2x \int d\rho \, L \,,
\ee
with the effective Lagrangian $L$ 
\ba \lb{8}
L & = & \frac{1}{2}\left[ \frac{m}{\mu_G}\,\zeta^2 
(\bm{X}\cdot(\dot{\bm{X}} \wedge \ddot{\bm{X}})) - m\,\zeta \
\dot{\bm{X}}^2 \right. \nonumber \\
& & \left. + \zeta\,\dot{\ol\psi}\,\bm{\Sigma} \cdot \bm{X}\,\dot{\psi} 
- \mu_E\,\ol{\psi}\,\dot{\psi} - 4\,\zeta^{-1}\Lambda \right] \,.
\ea
where $\cdot = \partial/\partial\rho$, the Minkowski scalar and wedge
products are defined by $\alpha \cdot \beta = \alpha^i \beta_i\,$, $(\alpha
\wedge \beta)_i = \varepsilon_{ijk} \alpha^j \beta^k$, the ``Dirac''
matrices $\Sigma^i$ are
\be \lb{Dirac}
\Sigma ^0 = \left(
\begin{array}{cc}
0 & 1 \\
-1 & 0
\end{array}
\right) \, , \,\,\,
\Sigma ^1 = \left(
\begin{array}{cc}
0 & -1 \\
-1 & 0
\end{array}
\right) \, , \,\,\,
\Sigma ^2 = \left(
\begin{array}{cc}
1 & 0 \\
0 & -1
\end{array}
\right) \, ,
\ee
and $\ol{\psi} \equiv \psi^T\,\Sigma^0$ is the (real) Dirac adjoint of the 
``spinor'' $\psi$. 

The action (\ref{7}) is invariant under the SL(2,R) $\approx$ SO(2,1)
group of transformations in the plane of the two Killing vectors $K^0$,
$K^1$. This invariance leads to the conservation of the ``angular
momentum'' vector
\be \lb{14}
\bm{J}=\bm{L}+\bm{S_G}+\bm{S_E} \,,
\ee
sum of ``orbital'' and ``spin'' contributions
\ba \lb{15}
\bm{L} & = & -m\,\bm{X} \wedge \dot{\bm{X}}\,, \nonumber\\
\bm{S_G} & = & -\frac{m}{2\mu_G}\, [\,2\bm{X}\wedge(
\bm{X}\wedge\ddot{\bm{X}})-\dot{\bm{X}} \wedge(
\bm{X}\wedge\dot{\bm{X}})\,]\,, \\
\bm{S_E} & = & \frac{1}{2}\, \Pi^T \!\bm{\Sigma} \psi \,, \nonumber
\ea
where $\Pi^T \equiv \partial L /\dot{\psi}$ is the moment canonically
conjugate to $\psi$, and we have set the Lagrange multiplier $\zeta$ equal
to 1. Variation of the Lagrangian (\ref{8}) with respect to
$\psi$ leads to the first integrals
\be \lb{PiT}
\Pi^T - \frac{\mu_E}{2} \ol{\psi} = 0
\ee
(we have without loss of generality chosen a gauge such that the constant
right-hand side is zero). The electromagnetic spin vector then reduces to
\be \lb{spin}
\bm{S_E} = \frac{\mu_E}{4}\,\ol{\psi} \bm{\Sigma} \psi \,.
\ee 
We note for future purposes that this vector is null,
\be \lb{null}
\bm{S}_E^{\,2} = 0 \,.
\ee

Varying the Lagrangian (\ref{8}) with respect to $\bm{X}$, and taking into
account equations (\ref{PiT}) and (\ref{spin}), we obtain the coupled
dynamical equations for the vector fields $\bm{X}$ and $\bm{S_E}$, 
\ba \lb{12}
\ddot{\bm{X}}  & = & -\ds\frac{1}{2\mu_G} [\,3(\dot{\bm{X}} \wedge
\ddot{\bm{X}}) +2(\bm{X} \wedge \dot{\ddot{\bm{X}}})\,] \nonumber\\
&   & +\ds\frac{2\mu_E}{mR^2} \bm{S_E} - \ds\frac{4\mu_E}{mR^4} \bm{X}
(\bm{S_E} \cdot \bm{X})\,,\\
\dot{\bm{S_E}}  & = &  \ds\frac{2\mu_E}{R^2} \bm{X}\wedge\bm{S_E}\,\, 
\Longleftrightarrow \,\,\dot{\psi} = \frac{\mu_E}{R^2}\, \bm{\Sigma} \cdot
\bm{X}\, \psi \nonumber
\ea
(for the choice $\zeta = 1$). Finally, the action (\ref{7}) is also
invariant under reparametrizations of $\rho$ (variations of $\zeta$),
leading to the Hamiltonian constraint,
\be \lb{Ham1}
H \equiv -\frac{m}{2} \dot{\bm{X}}^2+\frac{m}{\mu_G}(\bm{X} \cdot 
(\dot{\bm{X}} \wedge \ddot{\bm{X}})) +\frac{2\mu_E}{R^2}
\bm{S_E} \cdot \bm{X}+2m\Lambda = 0 
\ee
(where we have again taken $\zeta = 1$ after variation). Using the first 
equation (\ref{12}) to eliminate $\bm{S_E} \cdot \bm{X}$ from
(\ref{Ham1}), this may be rewritten in a form involving only the
gravitational field $\bm{X}$ and its derivatives, 
\be \lb{Ham2}
H \equiv -{m \over 2} \left[ \dot{\bm{X}}^2 + 2\bm{X} \cdot \ddot{\bm{X}}
+ {1 \over \mu_G} (\bm{X} \cdot (\dot{\bm{X}} \wedge \ddot{\bm{X}}) 
- 4\Lambda \right] = 0\,.
\ee
In the next two sections, we construct particular solutions to the coupled
equations (\ref{12}), with the integration constants constrained by 
(\ref{Ham1}).

\setcounter{equation}{0}
\section{Self-dual solutions}

In this section we search for electromagnetically self-dual solutions
\cite{Cam} to equations (\ref{12}), i.e. solutions such that,
\be \lb{sd}
\frac{1}{4}\,F_{\mu\nu}F^{\mu\nu} \equiv -\frac{2\mu_E}{R^2}\,\bm{S_E}
\cdot \bm{X} = 0\,.
\ee
It follows from equations (\ref{null}) and (\ref{sd}) that $\bm{X}$ is
spacelike, $\bm{X}^2 \equiv R^2 = - \sigma^2$ ($\sigma$ real), and that
the vector $\bm{S_E} \wedge \bm{X}$ is collinear to $\bm{S_E}$,
\be
\bm{S_E} \wedge \bm{X} = \pm \sigma \bm{S_E}\,.
\ee
This last equation reduces the second equation (\ref{12}) to 
\be \lb{lin}
\dot{\bm{S_E}}  = \pm \ds\frac{2\mu_E}{\sigma}\,\bm{S_E}\,,
\ee
which implies that the electromagnetic spin vector $\bm{S_E}$ has a
constant null direction $\bm{\beta}$ (it can be checked that $\bm{L}$ and
$\bm{S_G}$ are also aligned along this direction, which is that of the
total angular momentum $\bm{J}$). The vector $\bm{X}$ can be decomposed
along the two orthogonal directions $\bm{\beta}$ (lightlike) and $\bm{\alpha}$ 
(spacelike, with the convenient normalization $\bm{\alpha}^2 = - 1$) as
\be \lb{17}
\bm{X}(\rho) = \bm{\alpha}\,\sigma(\rho) + \bm{\beta}\,M(\rho)\,.
\ee
The $\bm{\alpha}$ component of the first equation (\ref{12}) gives 
$\ddot{\sigma} = 0$, which is solved by
\be \lb{22}
\sigma = a\rho + b\,.
\ee
The Hamiltonian constraint (\ref{Ham1}) then fixes the value of the
constant $a$,
\be
a^2 = - 4 \Lambda\,,
\ee 
so that these self-dual solutions exist only for $\Lambda \leq 0$.

In the case of a negative cosmological constant $\Lambda \equiv - l^{-2}$,
equation (\ref{lin}) integrates to
\be
\bm{S_E} = \mu_E k^2\,\rho^{\pm l\mu_E}\bm{\beta}\,,
\ee
where $k$ is an integration constant proportional to the electric charge
(we have chosen the origin of coordinates so that the constant $b$ in
(\ref{22}) vanishes). The $\bm{\beta}$ component of the first equation
(\ref{12}) then reduces to the ordinary differential equation
\be
2\rho\,\dot{\ddot{M}\,} + (3 \mp l\mu_G)\,\ddot{M} = \pm\,\frac{k^2 l^3 
\mu_E^2\,\mu_G}{8m}\,\rho^{\pm l\mu_E-2}\,,
\ee
which is solved by 
\be \lb{27}
M_{\pm}(\rho) = M_{\infty} + C \rho^{(1 \pm l\mu_G)/2}
+ D\,\rho^{\pm l\mu_E}\,, 
\ee
with
\be
D = \frac{k^2 l^2 \mu_E\,\mu_G}{2m(1 \mp l\mu_E)(1 \mp l(2\mu_E-\mu_G))}
\ee
(a possible linear term in (\ref{27}) can always be taken to zero by a
suitable redefinition of the constant vector $\bm{\alpha}$).
The parametrization \cite{part} of the vectors $\bm{\alpha}$ and $\bm{\beta}$,
\be \lb{25}
\bm{\alpha} = \frac{1}{2l}\,(1-l^2,\,1+l^2,\,0)\,, \qquad 
\bm{\beta} = -\frac{1}{4}(1+l^2,\,1-l^2,\,\mp 2l)\,,
\ee
then leads to the solution for the gravitational and electromagnetic fields,
\ba \lb{26}
ds^2 & = & \left( \frac{2\rho}{l^2} - \frac{M_{\pm}(\rho)}{2} \right) dt^2
\pm l M_{\pm}(\rho)\,d\theta\,dt -
\left( 2\rho + l^2 \frac{M_{\pm}(\rho)}{2} \right) d\theta^2 -
{l^2 d\rho^2 \over 4\rho^2}\,,\nonumber\\
A_{\mu}\,dx^{\mu} & = & k\,\rho^{\pm l\mu_E/2}\,(\,dt \mp l\,d\theta)\,,
\ea
where $\theta$ varies on the unit circle.
It is easily checked that this solution reduces to the self-dual solution
of pure TMG \cite{part} if $k = 0$, and to the self-dual solution of TME
coupled to Einstein gravity \cite{Cam} if $C = 0$ and $\mu_G \ra \infty$.
The metric (\ref{26}) is (as in the case of the self-dual solution to the
three-dimensional Einstein equations \cite{Cam}) horizonless, and is
asymptotic to the extreme BTZ metric\footnote{The BTZ radial coordinate
$r$ is related to our radial coordinate $\rho$ by $r^2 = 2 \rho$.} \cite{BTZ} 
with mass $M_{\infty} > 0$ and spin $J = \pm l M_{\infty}$ if
\be
\pm l \mu_E < 0 \qquad \mbox{and} \qquad \pm l \mu_G < -1\,,
\ee 
which we now assume (these inequalities imply that the two
Chern-Simons coupling constants $\mu_E$ and $\mu_G$ have the same sign).
The nature of the apparent singularity at $\rho = 0$ may be elucidated by
considering the first integral of the geodesic equation \cite{part}
\be \lb{geo}
\left( \frac{d\rho}{d\tau} \right)\!^2 + \bm{P} \cdot \bm{X} -
\varepsilon R^2 = 0\,,
\ee
where $\tau$ is an affine parameter, $\bm{P}$ a constant future lightlike 
vector, and
$\varepsilon = +1, 0$ or $-1$. The effective potential in (\ref{geo}) is
dominated near $\rho = 0$ by the term $\bm{\beta} \cdot \bm{P}\,M(\rho)$
which behaves as $-C\,\rho^{(1\pm l\mu_G)/2}$ if $\mp l\mu_G > 1 \mp
2l\mu_E$, and as $-(1/m)\rho^{\pm l\mu_E}$ if $\mp l\mu_G < 1 \mp
2l\mu_E$. It follows that if $C < 0$ in the first case, or $m < 0$ in the
second case, all the geodesics are reflected away from the singularity,
except for the spacelike geodesics ($\varepsilon = -1$) with $\bm{P} \cdot
\bm{X} = 0$; the circle $\rho = 0$ is at infinite affine distance
on these geodesics, so that the geometry and the electromagnetic field
(\ref{26}) are regular\footnote{In
the exceptional case $C = 0$ with $\mp l\mu_G > 1 \mp 2l\mu_E$, the geometry 
is regular for $m > 0$.}.

We now briefly consider the other case $\Lambda = 0$ ($a = 0$). Choosing
vectors $\bm{\alpha}$ and $\bm{\beta}$ of a form similar to (\ref{25}),
with $l$ replaced by $b$, we obtain the solution
\ba \lb{32}
ds^2 & = & \left( 1 - \frac{M_{\pm}(r)}{2} \right) dt^2
\pm b M_{\pm}(r)\,dtd\theta - b^2
\left( 1 + \frac{M_{\pm}(r)}{2} \right) d\theta^2 - dr^2\,, \nonumber \\
A_{\mu}\,dx^{\mu} & = & k\,\mbox{e}^{\pm \mu_E\,r}\,(dt \mp b\,d\theta)\, 
\ea
where we have put $\rho \equiv b\,r$, and the function $M_{\pm}(r)$ is given
by
\be
M_{\pm}(r) = Br + C\mbox{e}^{\pm \mu_G\,r} + D\mbox{e}^{\pm 2\mu_E\,r}\,,
\ee
with
\be
D = \frac{k^2 \mu_G}{2m(2\mu_E-\mu_G)}\,\mbox{e}^{\pm 2\mu_E\,r}\,.
\ee 
The radial coordinate in (\ref{32}) varies from $-\infty$ to $+\infty$. If
$\mu_E$ and $\mu_G$ are of the same sign, the metric (\ref{32}) is
asymptotically flat at one of the points at infinity, e.g. $r \ra +\infty$
for
\be
\pm \mu_E < 0 \qquad \mbox{and} \qquad \pm \mu_G < 0\,.
\ee  
The asymptotic metric is for $B = 0$ the cylindrical Minkowski spacetime (a
conical spacetime with the extremal value $2\pi$ for the deficit angle),
and for $B \neq 0$ the other extremal flat spacetime (equation (18) of
\cite{cosm}). The other point at infinity, $r \ra -\infty$, is at infinite
geodesic distance if $C < 0$ (or $C = 0$, $m > 0$) for $\mp \mu_G > \mp 2
\mu_E$, or if $m > 0$ for $\mp \mu_G < \mp 2\mu_E$. The full solution
(\ref{32}) is then perfectly regular.

\setcounter{equation}{0}
\section{Diagonal solutions}

Diagonal metrics, in the parametrization (\ref{5}), are such that
\be
Y = 0\,.
\ee
While the simplest solutions of Einstein gravity are diagonal, it is well
known \cite{Vuo} that the highly non-linear character of the equations of
sourceless TMG precludes the existence of diagonal solutions. However we
shall show that, quite remarkably, the (equally non linear) equations of
TMGE admit diagonal solutions in the case where the two Chern-Simons
coupling constants exactly balance, $\mu_G + \mu_E = 0$.

For our present purpose, it is appropriate to write the dynamical
equations (\ref{12}) in component notation for the components
\be \lb{35}
U \equiv T + X \,,\qquad V \equiv T - X \,,\qquad 
\xi \equiv A_0 \,,\qquad \eta \equiv A_1 \,.
\ee
For $Y = 0$, $U$ and $V$ are ``light-cone'' coordinates, with $R^2 = UV$.
With this parametrization, the second equation (\ref{12}) reads, for $Y = 0$, 
\be \lb{39}
\dot{\xi} = \mu_E\,{\eta \over V}\,,  \qquad
\dot{\eta} = - \mu_E\,{\xi \over U} 
\ee
(Maxwell-Chern-Simons equations), while the first equation (\ref{12}) 
reduces to
\ba \lb{40}
& & \ddot{U} = \frac{1}{m}\,{\dot{\xi}}^2\,, \qquad
\ddot{V} = \frac{1}{m}\,{\dot{\eta}}^2\,, \nonumber \\
& & \ddot{Y} = -\frac{1}{4\mu_G} \left[ 3(\dot{U}\ddot{V} - \ddot{U}\dot{V}) 
+ 2(U\dot{\ddot{V}}-\dot{\ddot{U}}V) \right] 
+ \frac{1}{m}\,\dot{\xi}\dot{\eta} = 0
\ea
(Einstein-Chern-Simons equations; the terms in factor of $1/m$ are the 
appropriate components of the Maxwell energy-momentum tensor).
The last equation (\ref{40}) may also be rewritten, by using equations 
(\ref{39})
to eliminate $U$ and $V$ and their first derivatives in terms of $\xi$,
$\eta$ and their derivatives, and the first two
equations (\ref{40}) to eliminate $\ddot{U}$ and $\ddot{V}$ and their
first derivatives in terms of $\dot{\xi}$, $\dot{\eta}$ and their
derivatives, as the second-order differential equation
\be \lb{42}
\xi \ddot{\eta} + \ddot{\xi} \eta + 2 \left( 3 + 2\,\frac{\mu_G}{\mu_E}
\right) \dot{\xi} \dot{\eta} = 0\,.
\ee
We shall also use the first integrals of the equations corresponding to
the conservation of the total angular momentum (\ref{14}),
\ba \lb{J}
& -{\ds\frac{m}{\mu_G}} \left[\,U(U\ddot{V} - \ddot{U}V) - \frac{1}{2}\dot{U}
(U\dot{V} - \dot{U}V)\,\right] - \mu_E\,\xi^2 & = 2J_u\,, \nonumber\\
& -{\ds\frac{m}{\mu_G}} \left[\,-V(U\ddot{V} - \ddot{U}V) + \frac{1}{2}\dot{V}
(U\dot{V} - \dot{U}V)\,\right] - \mu_E\,\eta^2 & = 2J_v\,, \\
& -m(U\dot{V} - \dot{U}V) - \mu_E\,\xi\eta & = 2J_y\,. \nonumber
\ea
Finally, the integration constants for our system 
must be constrained by the Hamiltonian constraint (\ref{Ham2}),
\be \lb{41}
U \ddot{V} + \dot{U} \dot{V} + \ddot{U} V - 4 \Lambda = 0 \,.
\ee

Although the system of equations (\ref{39}), (\ref{40}) is obviously
overdetermined, as there are five equations for only four unknown
functions, it is very difficult to prove that it does not (or does, for
special parameter values) admit solutions. The only case we have been able
to treat completely corresponds to the choice of integration constants
$\bm{J} = 0$ (the right-hand sides of the three equations (\ref{J}) are
zero). Then, it follows from the third equation (\ref{J}) that
\be \lb{Jy0}
U\dot{V} - \dot{U}V = -\frac{\mu_E}{m}\, \xi\eta \,.
\ee
Using this equation as well as equations (\ref{39}), we are able to show
that the first two equations (\ref{J}) integrate to
\be \lb{UV}
U = a\,\xi^2\eta^{2x}\,, \qquad
V = b\,\xi^{2x}\eta^2\,,
\ee
where we have put $x \equiv 1 + \mu_G/\mu_E$, and $a$, $b$ are integration
constants. These equations enable us to write the system (\ref{39}) as
\be \lb{49}
\dot{\xi} = {\ds\frac{\mu_E}{b}}\,\xi^{-2x}\eta^{-1} \,, \qquad
\dot{\eta} = -{\ds\frac{\mu_E}{a}}\,\xi^{-1}\eta^{-2x} \,, 
\ee
which enables us to reduce the crucial differential equation (\ref{42}) to
the algebraic relation
\be \lb{nogo}
x \left( 1 + \frac{a}{b} \left( \frac{\xi}{\eta} \right)^{-2x}\,
\right)^2 = 0 \,.
\ee

First assume $x \neq 0$ ($\mu_G + \mu_E \neq 0$). In this case equation
(\ref{nogo}) implies that the functions $\eta$ and $\xi$ are proportional,
so that (from equations (\ref{UV}) and (\ref{49}) $\dot{U}$ and $\dot{V}$
are constant, in contradiction with the first two equations (\ref{40}).
So equation (\ref{nogo}) can be satisfied only if $x = 0$, that is if the
two Chern-Simons coupling constants are related by
\be \lb{rel}
\mu_G + \mu_E = 0 \,.
\ee
Conversely we can show that the only diagonal solutions to the TMGE
equations with $\mu_G + \mu_E = 0$ are those with $\bm{J} = 0$. The
reasoning goes as follows. When the relation (\ref{rel}) holds, equation
(\ref{42}) reduces to $(\xi \eta)\,\ddot{ } = 0$, which is solved by 
\be \lb{50}
\xi \eta = c\rho + d \,.
\ee
We may then use equations (\ref{39}) to transform the last equation
(\ref{J}) into a first-order differential equation, depending on the
parameter $J_y$, for the unknown function $f \equiv \rho\,(\dot{\xi}/\xi -
\dot{\eta}/\eta)$. Using this equation, we are able to reduce the
Hamiltonian constraint (\ref{41}) to an algebraic relation between
$f(\rho)$ and $\rho$, which can be satisfied only if $J_y = 0$; the first
two equations (\ref{J}) then give $J_u = J_v = 0$ as well.

We now assume $x = 0$. A second differentiation then decouples the system
(\ref{49}) to
\be \lb{51}
\frac{\ddot{\xi}}{\dot{\xi}} - \frac{b}{a}\frac{\dot{\xi}}{\xi} = 0\,,\qquad
\frac{\ddot{\eta}}{\dot{\eta}} - \frac{a}{b}\frac{\dot{\eta}}{\eta} = 0\,.
\ee
We must now distinguish between two possibilities. If $b \neq a$
($c \neq 0$ in (\ref{50})), equations (\ref{51}) integrate to 
\be \lb{44}
\xi =\alpha \rho^p\,, \qquad \eta =\beta \rho^{1-p} \,,
\ee
with $p = a/(a - b)$. This leads, according to equations 
(\ref{39}), to
\be
U = - \frac{\alpha \mu_E}{\beta (1-p)}\,\rho^{2p}\,, \qquad
V = \frac{\beta \mu_E}{\alpha p}\,\rho^{2-2p} \,.
\ee
The first two equations (\ref{40}) are both satisfied provided the
three constants $\alpha$, $\beta$ and $p$ are related by
\be \lb{45}
\alpha \beta =2m\mu_E {\frac{(2p-1)}{p\,(p-1)}}\,.  
\ee
The Hamiltonian constraint (\ref{41}) then determines the possible values 
of the exponent $p$ to be
\be \lb{47}
p_{\pm } = \frac{1}{2} \left[ 1 \pm \sqrt{\frac{\Lambda +\mu_E^2}
{\Lambda -\mu_E^2}} \right] 
\ee
for $\Lambda^2 > \mu_E^4\,$. Using the relation $p_+ + p_- = 1$, the full
solution may be written, for $p = p_-$,
\ba \lb{48}
ds^2 & = & 4 \varepsilon  m\sqrt{\Lambda^2 - \mu_E^4}
\left(  {\rho^{2p_-} \over \beta^2 p_+} ({dx^0})^2 - 
{\rho^{2p_+} \over \alpha^2 p_-} ({dx^1})^2  \right) +
{d\rho^2 \over 2(\Lambda - \mu_E^2) \rho^2} \,, \nonumber \\
A_{\mu}\,dx^{\mu} & = & \alpha \rho^{p_-}\,dx^0 + \beta
\rho^{p_+}\,dx^1 \,,
\ea
where $\varepsilon = {\rm sign}\Lambda$, and the two constants $\alpha$
and $\beta$ are constrained by (\ref{45}) (the other choice $p = p_+$ leads
to the same solution with the irrelevant exchange of coordinate labels 
$ 0 \leftrightarrow 1$). 

Let us discuss the geometrical properties of this diagonal solution. 
If $\underline{\Lambda > \mu_E^2}$ ($\varepsilon > 0$), 
the exponents $p_+$ and $p_-$ have opposite signs ($p_+ > 1$, $p_- <0$), 
so that the solution (\ref{48}) corresponds to a Riemannian spacetime
for $m > 0$, or to an anisotropic cosmology (with $\rho$ as time
coordinate, $x^0$ and $x^1$ as space coordinates) for $m < 0$. If
$\underline{\Lambda < - \mu_E^2}$ ($\varepsilon < 0$), $p_+$ and $p_-$ are
both positive, so that the metric (\ref{48}) has the Lorentzian
signature,with $t = x^1$ for $m > 0$ or $t = x^0$ for $m < 0$. The static
rotationally symmetric solution corresponding to the local solution
(\ref{48}) is thus
\ba \lb{stat}
ds^2 & = & 4|m|\sqrt{\Lambda^2 - \mu_E^4}
\left( \frac{\rho^{2p_{\pm}}}{\alpha_{\mp}^2 p_{\mp}}\,dt^2 - 
\frac{\rho^{2p_{\mp}}}{\alpha_{\pm}^2 p_{\pm}}\,d\theta^2 \right) 
- \frac{d\rho^2}{2(\mu_E^2 - \Lambda) \rho^2} \,, \nonumber \\
A_{\mu}\,dx^{\mu} & = & \alpha_{\pm} \rho^{p_{\pm}}\,dt 
+ \alpha_{\mp} \rho^{p_{\mp}}\,d\theta \,,
\ea 
where we have put $\alpha = \alpha_-$, $\beta = \alpha_+$, and
$\pm = {\rm sign}(m)$. The study of the geodesic equation (\ref{geo}) for
this case shows that the metric (\ref{stat}) is, for both signs of $m$,
singular at the point $\rho = 0$. All geodesics terminate at $\rho = 0$
for $m > 0$, while for $m < 0$ (the preferred sign in TMGE) only radial
geodesics terminate at $\rho = 0$. This last property suggests that the
$m < 0$ static solution (\ref{stat}) is the TMGE analogue of the conical 
point particle solution of three-dimensional Einstein gravity. Indeed, the
extreme black-hole solutions of dilaton gravity with $a > 1$
\cite{dilaton}, which exhibit a similar property (only radial geodesics
reach the pointlike horizon) have been shown to also behave as elementary
particles in other respects \cite{HW} (another ``black point'' metric is
discussed in \cite{Soleng}). We also note that the $\bm{J} = 0$ solution 
(\ref{stat}) reduces in the limit $-\Lambda \equiv
l^2 \ra \mu_E^2$, $m \ra \infty$ ($G \ra 0$) to the $\bm{L} = 0$ BTZ
``vacuum'' solution \cite{BTZ}
\be
ds^2 = \frac{2\rho}{l^2}\,dt^2 - 2\rho\,d\theta^2 
- \frac{l^2\,d\rho^2}{4\rho^2}\,.
\ee

For the other possibility $b = a$ ($c = 0$ in (\ref{50}), equations
(\ref{51}) integrate to 
\be
\xi = k\,{\rm e}^{\alpha \rho}\,, \qquad \eta = hk\,{\rm e}^{-\alpha \rho}\,.
\ee
We then find that equations (\ref{39})--(\ref{41}) are satisfied for
$\Lambda = \mu_E^2$ if the constants $h$, $k$, $\alpha$ are related by $h
= 4m\mu_E/k^2 \alpha$. Redefining the coordinates $x^0 \equiv x$,
$hx^1 \equiv y$, $\alpha \rho \equiv \mu_E t$, we obtain the
$\underline{\Lambda = \mu_E^2}$ diagonal solution
\ba
ds^2 & = & \frac{k^2}{4m} \left( {\rm e}^{2\mu_E t}\,dx^2 
+ {\rm e}^{-2\mu_E t}\,dy^2 \right) + dt^2 \,, \nonumber\\
A_{\mu}\,dx^{\mu} & = & k \left( {\rm e}^{\mu_E t}\,dx 
+ {\rm e}^{-\mu_E t}\,dy \right)\,,
\ea
corresponding to a Riemannian spacetime for $m > 0$, or to an anisotropic
area-preserving cosmology for $m < 0$.

\section{Conclusion}

Despite the apparent complexity of the coupled field equations of 
topologically massive electrodynamics, we have been able to construct two
families of exact solutions to the theory with two Killing vectors. The
self-dual stationary solutions of Section 3 generalize similar solutions
of TMG or of TME coupled to Einstein gravity. These solutions, which exist
only for $\Lambda \leq 0$, are asymptotic to extreme BTZ metrics if the
Chern-Simons coupling constants $\mu_G$ and $\mu_E$ have the same sign,
with $\Lambda > -\mu_G^2$,
and are regular for suitable ranges of the model parameters. The diagonal
solutions which we have obtained in Section 4 in the case where the two
Chern-Simons coupling constants exactly balance, $\mu_G + \mu_E = 0$, 
include anisotropic
cosmologies if $\Lambda \geq \mu_E^2$ and $G < 0$, and static solutions 
with a pointlike horizon if $\Lambda < -\mu_E^2$ and $G < 0$. 

Although we have
not been able to prove it, it seems very likely that our system does not
admit static solutions for $\mu_G + \mu_E \neq 0$ (it certainly does not 
admit any either for $\mu_E = 0$ or for $\mu_G = 0$). So we conjecture that 
our solutions (\ref{stat}) are in fact the unique static rotationally 
symmetric solutions to TMGE. As in the case of other models 
\cite{dilaton}\cite{Soleng}, we expect that
these solutions with pointlike horizons are the limit of solutions with
regular horizons to the $\mu_G + \mu_E = 0$ theory. Such hypothetical
black-hole solutions would necessarily be non-static.

\end{document}